# Adiabatic and Nonadiabatic Energy Dissipation during Scattering of Vibrationally Excited CO from Au(111)

Meng Huang,[1] Xueyao Zhou,[1,2] Yaolong Zhang,[2] Linsen Zhou,[1] Maite Alducin,[3,4] Bin Jiang,[2] and Hua Guo[1,5,*]

[1]Department of Chemistry and Chemical Biology, University of New Mexico, Albuquerque, New Mexico 87131, USA
[2]Department of Chemical Physics, University of Science and Technology of China, Hefei, Anhui 230026, China
[3]Centro de Física de Materiales, CSIC-UPV/EHU, P. Manuel de Lardizabal 5, 20018 San Sebastián, Spain
[4]Donostia International Physics Center DIPC, P. Manuel de Lardizabal 4, 20018 San Sebastián, Spain
[5]Department of Physics and Astronomy, University of New Mexico, Albuquerque, New Mexico 87131, USA

A high-dimensional potential energy surface (PES) for CO interaction with the Au(111) surface is developed using a machine-learning algorithm. Including both molecular and surface coordinates, this PES enables the simulation of the recent experiment on scattering of vibrationally excited CO from Au(111). Trapping in a physisorption well is observed to increase with decreasing incidence energy. While energy dissipation of physisorbed CO is slow, due to weak coupling with both the phonons and electron-hole pairs, its access to the chemisorption well facilitates fast vibrational relaxation of CO through nonadiabatic coupling with surface electron-hole pairs.

Energy transfer between molecules and metal surfaces represents a key aspect of surface processes, with important implications in a wide array of interfacial phenomena. There are two major energy exchange channels, namely the adiabatic coupling with surface phonons and the nonadiabatic interaction with electron-hole pairs (EHPs).[1-3] The lifetime of CO($v$=1) adsorbate has been measured to be 1-2 ps on Cu(001), using several experimental techniques.[4-7] Such a short lifetime for a high frequency mode ($\omega$=2129 cm$^{-1}$) can only be explained by its nonadiabatic coupling with surface EHPs, because its direct coupling with the low-frequency phonons is unlikely. This nonadiabatic energy dissipation mechanism has been characterized by various theoretical models,[8-18] cumulating with the latest first-principles calculations that quantitatively reproduced the observed lifetime.[19,20]

It was thus a surprise when Shirhatti et al. reported a long lifetime (~10$^2$ ps) for trapped CO($v$=1) in the scattering of vibrationally excited CO($v$=2) from Au(111).[21] It was postulated that physisorption might be involved, given the relatively low desorption temperature of CO from Au(111).[22] Indeed, a recent density functional theory (DFT) study by Lončarić et al. did find such a physisorption well for CO on Au(111),[23] using the Bayesian Error Estimation Functional method with van der Waals corrections (BEEF-vdW).[24] The lifetime of physisorbed CO($v$=1) was calculated within first-principles many-body perturbation theory and found to be consistent with the experimental value.[21] The long vibrational lifetime was attributed to the weaker couplings with EHPs because of the large distance between the adsorbate and surface. The same argument has also been used to explain the vibrationally hot precursor CH$_4$ on the Ir(111) surface.[25]

However, the aforementioned theoretical work was only intended to calculate the vibrational relaxation rate for CO adsorbed on the surface, and it provides information on neither the mechanism and dynamics on how the impinging CO molecules are trapped and then desorbed, nor the accompanying energy dissipation into surface phonons. In principle, Ab Initio Molecular Dynamics (AIMD) can shed light on such issues, but the trapping and diffusion are too rare and too long to be computationally feasible for the on-the-fly method. To meet this challenge, we report here a machine-learning approach which trains neural networks (NNs) to predict the high-dimensional potential energy surface (PES) for the CO/Au(111) system, thus avoiding the expensive on-the-fly DFT calculations in AIMD. Based on the original idea of Behler and Parrinello,[26,27] atomistic NNs (AtNN) can be designed to include both the molecular and surface degrees of freedom (DOFs) within a periodic slab model,[28-31] thus allowing adiabatic energy exchange between the impinging molecule and surface phonons. To this end, a 60-dimensional PES is trained using both energies and gradients from AIMD calculations, which enables large numbers of quasi-classical trajectories (QCTs) to determine trapping probabilities and to follow the long-time diffusion dynamics of the trapped species. In addition, a generalized Langevin equation (GLE) with approximate friction coefficients is used to simulate the nonadiabatic energy dissipation to surface EHPs.[2] A combination of these theoretical advances allows a detailed characterization of the trapping and energy dissipation during the scattering, thus shedding valuable light on the intricate interplay between adiabatic and nonadiabatic energy exchanges.

To generate the initial data points for building the PES, AIMD simulations of CO scattering from Au(111) were first performed using the Vienna Ab Initio Simulation Package (VASP) [32,33] with the BEEF-vdW functional.[24] In these simulations, Au(111) was approximated by a slab with 4 layers of a 3×3 unit cell with the bottom two layers frozen, which is separated from its images by 16 Å of vacuum. The cut-off energy in the planewave basis was 450 eV and the Brillouin zone sampled with a 5×5×1 Monkhorst-Pack mesh. The slab was thermalized to 300K and the geometries and velocities of the surface atoms were randomly sampled. The CO($v$=2, $J$=0) molecule, with its internal coordinate and momentum sampled on a CO potential, was prepared from 8 Å above the surface with random orientations and positions in the unit cell. Following the experiment,[21] the incidence angle was fixed at $\theta$=9° from the surface normal. A total of 100 and 80 trajectories was calculated respectively at the experimental incidence energies of 0.64 and 1.28 eV.[21]

In the Behler−Parrinello approach,[26] the total energy of the system is obtained by summing atomic energies, which are represented by AtNNs for different atomic types. The environment of an atom is described by symmetry functions, which contain two- and three-body interactions.[27,34] The NN was trained by 10766 DFT points with the root mean square errors in energy and force of 9.78 meV and 20.00 meV/Å. More details of the fitting are given in Supporting Information (SI).

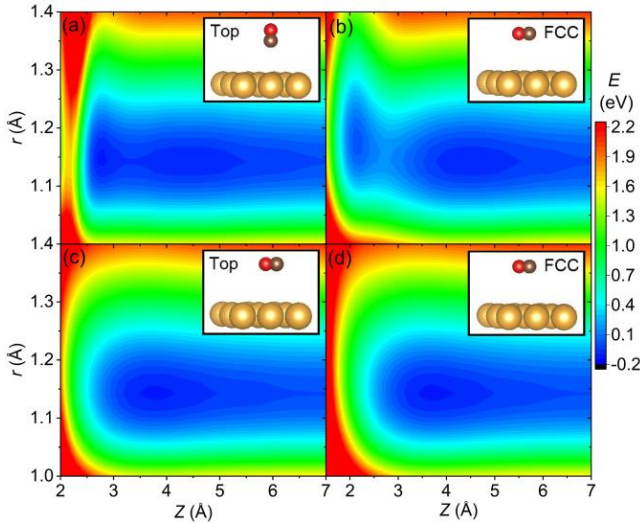

FIG. 1 Two-dimensional PES cuts at the top (a,c) and fcc (b,d) sites with different CO orientations.

Two-dimensional cuts of the AtNN PES at two surface sites are displayed in Fig. 1 as functions of the CO bond length ($r$) and distance from the CO center of mass (COM) to surface ($Z$), with CO oriented either parallel and perpendicular to the surface normal, and the surface atoms were kept frozen at their equilibrium positions. Similar to the previous theoretical work,[23] there is a physisorption well with a depth of ~0.10 eV and a large distance from the surface ($Z$~4.0 Å). The parallelly oriented CO has a deeper well than the perpendicularly oriented CO. In addition, a chemisorption well with 0.13 eV in depth is found at the top site, while it becomes metastable at the hollow site. These chemisorption wells feature perpendicularly oriented CO with a shorter C-O bond and are much closer to the surface ($Z$~2.9 Å). A barrier of 61 meV exists from the physisorption well. The adsorption energies in various wells are somewhat smaller than the experimental estimation (0.18±0.10 eV).[22]

The dynamics on the AtNN PES is more than $10^4$ times faster than on-the-fly AIMD, thus enabling many more trajectories with much longer propagation time. To explore the scattering dynamics, a total of 15,000 QCT trajectories was launched towards the surface at a surface temperature $T_s$=300K with the incident kinetic energy $E_{in}$ set at the experimental values, 0.64, 0.40 and 0.32 eV. The other initial conditions are identical to the AIMD calculations, but the propagation time is extended to 50 ps. More details of QCT calculations can be found in SI.

While the majority of the trajectories undergo scattering back to the vacuum, there is a small portion that never desorbs at the end of the 50 ps run, which are denoted as "trapped" (T). The scattered trajectories are further divided into two categories; the ones with a single inner turning point are classified as "direct scattered" (DS), while those with multiple inner turning points are called "trapped then scattered" (TS). The fractions of these three types of trajectories are given in Table S1 in SI, along with the averaged translational ($<E_{trans}>$), rotational ($<E_{rot}>$), and vibrational ($<E_{vib,f}>$) energies. The probability of the trapped trajectories is slightly over 0.2% at 0.64 eV, but increases to ~ 3% at 0.32 eV. This agrees with the observed experimental trend that significantly long-lived CO molecules were found at low incidence energies.[21]

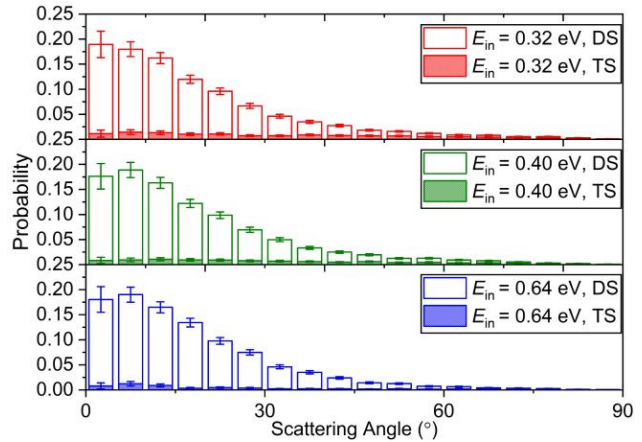

FIG. 2 Angular distributions of the scattered CO for DS (shaded) and TS (solid) trajectories at three incidence energies.

The angular distributions of the scattered CO are displayed in **Error! Reference source not found.**. As expected, the dominant DS trajectories are mostly specular, with the distribution mostly centered around 10°, in good agreement with the experimental value of ~9°.[21] On the other hand, TS trajectories have a much broader angular distribution, also consistent with the experiment.[21]

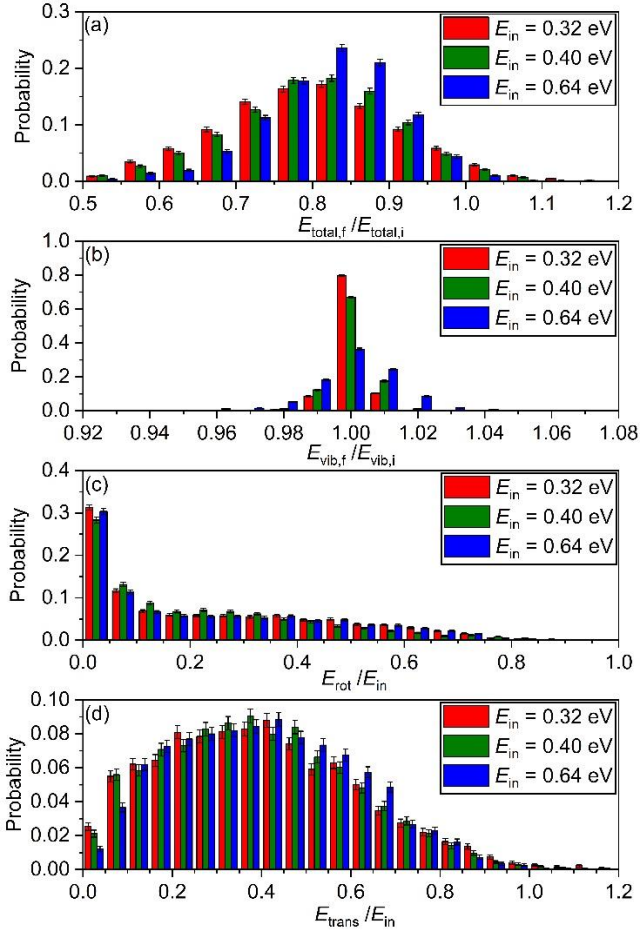

FIG. 4 Distributions of the total energy ratio ($E_{total,f}/E_{total,i}$), vibrational energy ratio ($E_{vib,f}/E_{vib,i}$), relative rotational energy ratio ($E_{rot}/E_{in}$), relative translational energy ratio ($E_{trans}/E_{in}$) for scattered CO at three different incidence energies.

As shown in Table S1, the collision of CO with the surface results in energy redistribution among different DOFs. Figure 3 displays ratios between the final and initial energies in different DOFs. As seen in Fig. 3(a), the total energy of CO decreases about 20% after scattering, apparently lost to surface DOFs. Among the molecular DOFs, the vibrational energy ratio (Fig. 3(b)) ranges from 0.90 to 1.10, suggesting strong vibrational elasticity, consistent with the large frequency mismatch between CO vibration and surface phonons. On the other hand, significant energy is transferred to CO rotation, as shown in Fig. 3(c). Figure 3(d) indicates that the energy loss in the translational DOF is quite substantial. Beyond the limit of $E_{trans}/E_{in}=0$ is the trapping of CO, in which the molecule has insufficient kinetic energy to escape the adsorption well.

Trapping probabilities for the three incident energies are shown in Fig. 4(a) to decay exponentially and the lifetimes have been estimated. The lifetime ($\tau$) of trapped CO on the surface, extracted from the slope in the logarithmic plot, are 18.4, 26.7, and 32.9 ps for $E_{in}$ = 0.64, 0.40, and 0.32 eV, respectively. While qualitatively similar, these lifetimes are quantitatively shorter than those reported in the experiment (~100 ps).[21] There might be several possible reasons for the underestimation. One possibility is the involvement of EHPs, which can in principle further dissipate the energy of trapped CO and extend its trapping lifetime.

To examine this possibility, we include EHPs using the GLE approach,[35] treating the nonadiabatic energy dissipation as electronic friction. In particular, the atomic friction coefficients of CO were obtained within the local density friction approximation (LDFA),[36] in which CO is assumed to move in a free-electron gas at the metal surface. The surface electron density was approximated from DFT calculations of perfect Au(111), which is expected to be a reasonable approximation even for the moving surface.[37] The friction coefficients of the C and O atoms in the electron gas, which are proportional to the transport cross-section at the Fermi level,[38] can be calculated with the position of the atom and the corresponding electron density (see SI for more details). The GLE dynamics were carried out with the same initial conditions, and the nonadiabatic trapping probabilities shown in Fig. 4(b) are not much different from the adiabatic results. Neither are other dynamic attributes shown in SI. This can be readily understood as the electron density in the physisorption well is vanishingly small, resulting in negligible friction coefficients. The weak EHP coupling is consistent with the long lifetime reported in the recent experiment[21] and first-principles calculations of the vibrational relaxation of CO physisorbed on Au(111).[23]

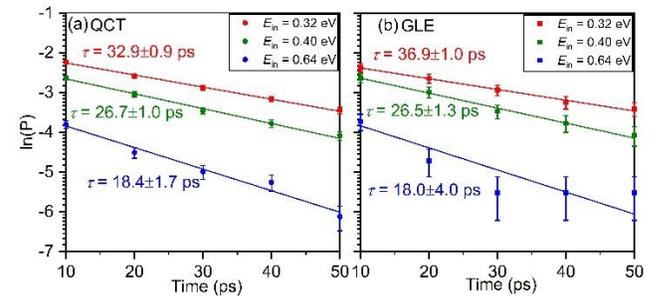

FIG. 3 Trapping probabilities of the CO as a function of time for (a) QCT trajectories within adiabatic approximation (b) GLE trajectories with electronic friction.

The discrepancy between the measured and calculated lifetime might thus have other origins. First, the experimental lifetime is merely an estimation from the time-of-flight of desorbed CO, which thus contain large uncertainties.[21] Theoretically, the calculated well depth (~0.1 eV) is shallower than the experimental estimate of 0.18 eV,[22] which could lead to a substantial reduction of lifetime. Further, the energy gained by the surface phonons from the impinging CO cannot dissipate as our simulations were performed in a microcanonical ensemble, and as a result, it might promote desorption and underestimate the trapping probability, especially after long time.[39] Finally, LDFA might not be sufficiently accurate, as suggested by the

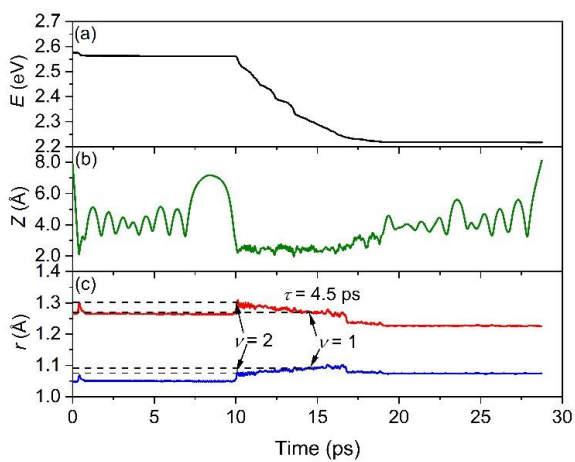

FIG. 5 Evolution of (a) total energy of the system (b) CO COM distance from surface (Z) (c) CO bond length (r) at the outer (red) and inner (blue) vibrational turning points of an exemplary trajectory at $E_{in}$ = 0.32 eV

alternative first-principles friction coefficients.[14,40,41]

A remaining puzzle is the CO($v$=1) product observed in the experiment, which has a fast and slow components.[21] As discussed above, our adiabatic simulations produce little vibrational relaxation and the only possible mechanism is due to EHPs. Yet, our results above also indicate that this nonadiabatic mechanism is unlikely if the CO is in the physisorption well. Interestingly, a few GLE trajectories are found to explore the chemisorption well and loss a large amount of energy, due to much larger friction coefficients stemming from much higher electron density near the surface. One such trajectory is scattered directly while others undergo trapping in the physisorption well. In Fig. 5, the energy loss of such an exemplary trajectory is shown, along with the $Z$ and $r$ coordinates. It is clear that the energy dissipation occurring in the chemisorption well leads to significant energy loss in the vibrational DOF. It is thus conceivable that the experimentally observed trapped CO($v$=1) stems from initial access of the chemisorption well, where rapid vibrational relaxation takes place. The direct scattering of such relaxed molecules leads to the fast component of the velocity distribution of desorbed CO($v$=1) while those trapped in the physisorption well give rise to the slow component.

To conclude, a high-dimensional PES developed with a machine-learning algorithm allows detailed simulations of CO scattering from Au(111). Calculated attributes of the scattered trajectories, such as the angular distributions, are in excellent agreement with experimental observations. Non-negligible trapping in the physisorption well is observed after the impinging molecule loses its incidence energy to surface phonons and other molecular DOFs. Because of the large separation between the physisorbed molecule and the surface, EHPs play a minor role in vibrational relaxation. However, it is shown that facile energy loss in the vibrational DOF is enabled by the access to the chemisorption well. Hence, the experimentally observed CO($v$=1) product is attributable to the rapid nonadiabatic vibrational relaxation in the chemisorption well by EHPs, in which the directly scattered CO constitutes the fast component of the velocity distribution while the subsequent trapping in the physisorption well is responsible for the slow component. The results presented here provided valuable insights into the experimental observations and have important implications in gas-metal interactions in general, particularly on the possible vibrational enhancement of reactivity for precursor-mediated surface reactions.

We acknowledge support by US National Science Foundation (CHE-1462109 to H.G.), National Key R&D Program of China (2017YFA0303500 to B.J.), National Natural Science Foundation of China (21573203, 91645202, and 21722306 to B.J.), Anhui Initiative in Quantum Information Technologies (AHY090200 to B.J.), and the Spanish Ministerio de Economía, Industria y Competitividad (FIS2016-76471-P to M.A.).

[1] J. C. Tully, Annu. Rev. Phys. Chem. **51**, 153 (2000).
[2] M. Alducin, R. Díez Muiño, and J. I. Juaristi, Prog. Surf. Sci. **92**, 317 (2017).
[3] S. P. Rittmeyer, V. J. Bukas, and K. Reuter, Adv. Phys. X **3**, 1381574 (2018).
[4] R. Ryberg, Phys. Rev. B **32**, 2671 (1985).
[5] J. D. Beckerle, R. R. Cavanagh, M. P. Casassa, E. J. Heilweil, and J. C. Stephenson, J. Chem. Phys. **95**, 5403 (1991).
[6] M. Morin, N. J. Levinos, and A. L. Harris, J. Chem. Phys. **96**, 3950 (1992).
[7] T. A. Germer, J. C. Stephenson, E. J. Heilweil, and R. R. Cavanagh, J. Chem. Phys. **101**, 1704 (1994).
[8] B. N. J. Persson and M. Persson, Solid State Commun. **36**, 175 (1980).
[9] B. Hellsing and M. Persson, Phys. Scr. **29**, 360 (1984).
[10] T. T. Rantala and A. Rosén, Phys. Rev. B **34**, 837 (1986).

[11] M. Head-Gordon and J. C. Tully, J. Chem. Phys. **96**, 3939 (1992).
[12] V. Krishna and J. C. Tully, J. Chem. Phys. **125**, 054706 (2006).
[13] M. Forsblom and M. Persson, J. Chem. Phys. **127**, 154303 (2007).
[14] M. Askerka, R. J. Maurer, V. S. Batista, and J. C. Tully, Phys. Rev. Lett. **116**, 217601 (2016).
[15] R. J. Maurer, M. Askerka, V. S. Batista, and J. C. Tully, Phys. Rev. B **94**, 115432 (2016).
[16] D. Novko, M. Alducin, M. Blanco-Rey, and J. I. Juaristi, Phys. Rev. B **94**, 224306 (2016).
[17] R. Scholz, G. Floß, P. Saalfrank, G. Füchsel, I. Lončarić, and J. I. Juaristi, Phys. Rev. B **94**, 165447 (2016).
[18] S. P. Rittmeyer, J. Meyer, and K. Reuter, Phys. Rev. Lett. **119**, 176808 (2017).
[19] D. Novko, M. Alducin, and J. I. Juaristi, Phys. Rev. Lett. **120**, 156804 (2018).
[20] D. Novko, J. C. Tremblay, M. Alducin, and J. I. Juaristi, Phys. Rev. Lett. **122**, 016806 (2019).
[21] P. R. Shirhatti *et al.*, Nat. Chem. **10**, 592 (2018).
[22] D. P. Engelhart, R. J. V. Wagner, A. Meling, A. M. Wodtke, and T. Schäfer, Surf. Sci. **650**, 11 (2016).
[23] I. Lončarić, M. Alducin, J. I. Juaristi, and D. Novko, J. Phys. Chem. Lett. **10**, 1043 (2019).
[24] J. Wellendorff, K. T. Lundgaard, A. Møgelhøj, V. Petzold, D. D. Landis, J. K. Nørskov, T. Bligaard, and K. W. Jacobsen, Phys. Rev. B **85**, 235149 (2012).
[25] E. Dombrowski, E. Peterson, D. Del Sesto, and A. L. Utz, Catal. Today **244**, 10 (2015).
[26] J. Behler and M. Parrinello, Phys. Rev. Lett. **98**, 146401 (2007).
[27] J. Behler, J. Chem. Phys. **134**, 074106 (2011).
[28] B. Kolb, X. Luo, X. Zhou, B. Jiang, and H. Guo, J. Phys. Chem. Lett. **8**, 666 (2017).
[29] Q. Liu, X. Zhou, L. Zhou, Y. Zhang, X. Luo, H. Guo, and B. Jiang, J. Phys. Chem. C **122**, 1761 (2018).
[30] K. Shakouri, J. Behler, J. Meyer, and G.-J. Kroes, J. Phys. Chem. Lett. **8**, 2131 (2017).
[31] Y. Zhang, X. Zhou, and B. Jiang, J. Phys. Chem. Lett. **10**, 1185 (2019).
[32] G. Kresse and J. Furthmuller, Phys. Rev. B **54**, 11169 (1996).
[33] G. Kresse and J. Furthmuller, Comp. Mater. Sci. **6**, 15 (1996).
[34] J. S. Smith, O. Isayev, and A. E. Roitberg, Chem. Sci. **8**, 3192 (2017).
[35] M. Head-Gordon and J. C. Tully, J. Chem. Phys. **103**, 10137 (1995).
[36] J. I. Juaristi, M. Alducin, R. Díez Muiño, H. F. Busnengo, and A. Salin, Phys. Rev. Lett. **100**, 116102 (2008).
[37] D. Novko, M. Blanco-Rey, M. Alducin, and J. I. Juaristi, Phys. Rev. B **93**, 245435 (2016).
[38] M. J. Puska and R. M. Nieminen, Phys. Rev. B **27**, 6121 (1983).
[39] J. Meyer and K. Reuter, Angew. Chem. Int. Ed. **53**, 4721 (2014).
[40] R. J. Maurer, B. Jiang, H. Guo, and J. C. Tully, Phys. Rev. Lett. **118**, 256001 (2017).
[41] P. Spiering and J. Meyer, J. Phys. Chem. Lett. **9**, 1803 (2018).